\documentclass[
 reprint,
superscriptaddress,
 amsmath,amssymb,
 aps,
superscriptaddress,
 amsmath,amssymb,
 aps,longbibliography
]{revtex4-1}

\usepackage{graphicx}
\usepackage{dcolumn}
\usepackage{bm}
\usepackage{amsmath,amssymb,amstext,amsthm}
\usepackage{braket}
\usepackage{bbold}
\usepackage{bbm}
\usepackage{xcolor}
\usepackage{ gensymb }
\usepackage{soul}
\setstcolor{red}

\begin{document}

\preprint{APS/123-QED}
\setlength{\abovedisplayskip}{1pt}
\title{Phase and contrast moir\'e signatures in two-dimensional cone beam interferometry}

\author{D. Sarenac}
\email{dsarenac@uwaterloo.ca}
\affiliation{Institute for Quantum Computing, University of Waterloo,  Waterloo, ON, Canada, N2L3G1}
\affiliation{Department of Physics, University at Buffalo, State University of New York, Buffalo, New York 14260, USA}

\author{G. Gorbet} 
\affiliation{Department of Physics, University of Waterloo, Waterloo, ON, Canada, N2L3G1}

\author{Charles W. Clark}
\affiliation{Joint Quantum Institute, National Institute of Standards and Technology and University of Maryland, College Park, Maryland 20742, USA}

\author{D. G. Cory}
\affiliation{Institute for Quantum Computing, University of Waterloo,  Waterloo, ON, Canada, N2L3G1}
\affiliation{Department of Chemistry, University of Waterloo, Waterloo, ON, Canada, N2L3G1}

\author{H. Ekinci} 
\affiliation{Institute for Quantum Computing, University of Waterloo,  Waterloo, ON, Canada, N2L3G1}

\author{M. E. Henderson} 
\affiliation{Institute for Quantum Computing, University of Waterloo,  Waterloo, ON, Canada, N2L3G1}
\affiliation{Department of Physics, University of Waterloo, Waterloo, ON, Canada, N2L3G1}

\author{M. G. Huber}
\affiliation{National Institute of Standards and Technology, Gaithersburg, Maryland 20899, USA}

\author{D. Hussey}
\affiliation{National Institute of Standards and Technology, Gaithersburg, Maryland 20899, USA}

\author{C. Kapahi} 
\affiliation{Institute for Quantum Computing, University of Waterloo,  Waterloo, ON, Canada, N2L3G1}
\affiliation{Department of Physics, University of Waterloo, Waterloo, ON, Canada, N2L3G1}

\author{P. A. Kienzle}
\affiliation{National Institute of Standards and Technology, Gaithersburg, Maryland 20899, USA}

\author{Y. Kim}
\affiliation{National Institute of Standards and Technology, Gaithersburg, Maryland 20899, USA}
\affiliation{University of Maryland, College Park, Maryland 20742, USA}

\author{M. A. Long} 
\affiliation{Materials Science and Technology Division, Los Alamos National Laboratory, Los Alamos, NM 87545, USA}

\author{J. D. Parker} 
\affiliation{J-PARC Center, Japan Atomic Energy Agency (JAEA), 2-4 Shirakata, Tokai, Ibaraki 319-1195, Japan}

\author{T. Shinohara} 
\affiliation{J-PARC Center, Japan Atomic Energy Agency (JAEA), 2-4 Shirakata, Tokai, Ibaraki 319-1195, Japan}

\author{F. Song} 
\affiliation{J-PARC Center, Japan Atomic Energy Agency (JAEA), 2-4 Shirakata, Tokai, Ibaraki 319-1195, Japan}

\author{D. A. Pushin}
\email{dmitry.pushin@uwaterloo.ca}
\affiliation{Institute for Quantum Computing, University of Waterloo,  Waterloo, ON, Canada, N2L3G1}
\affiliation{Department of Physics, University of Waterloo, Waterloo, ON, Canada, N2L3G1}

\date{\today}


\pacs{Valid PACS appear here}


\begin{abstract}
Neutron interferometry has played a distinctive role in fundamental science and characterization of materials. Moir\'e neutron interferometers are candidate next-generation instruments: they offer microscopy-like magnification of the signal, enabling direct camera recording of interference patterns across the full neutron wavelength spectrum. Here we demonstrate the extension of phase-grating moir\'e interferometry to two-dimensional geometries. Our fork-dislocation phase gratings reveal phase singularities in the moir\'e pattern, and we explore orthogonal moir\'e patterns with two-dimensional phase-gratings. Our measurements of phase topologies and gravitationally induced phase shifts are in good agreement with theory. These techniques can be implemented in existing neutron instruments to advance interferometric analyses of emerging materials and precision measurements of fundamental constants.
\end{abstract}
\maketitle

\section{\label{sec:level1}Introduction}

Perfect-crystal neutron interferometry possesses a prestigious record of high impact fundamental science experiments such as the observation of gravitationally induced quantum interference~\cite{colella1975observation}, 4$\pi$ symmetry of spinor rotation~\cite{rauch1975verification}, neutron triply-entangled GHZ states and quantum discord ~\cite{hasegawa2010engineering,wood2014quantum}, matter-wave orbital angular momentum~\cite{clark2015controlling,sarenac2016holography}, and the probing of dark energy and fifth forces~\cite{lemmel2015neutron,li2016neutron}. This is in-part due to the unique properties of the neutron such as it's electrical neutrality, relatively large mass, and angstrom sized wavelengths~\cite{klein1983neutron,abele2008neutron,willis2017experimental}. Such properties also make the neutron a convenient and indispensable probe of modern materials as they are capable of characterizing bulk properties and nanometer-sized spin textures~\cite{fuhrman2015interaction,qian2018new,kezsmarki2015neel,henderson2023three}.

A recent focus in neutron interferometry has been in grating-based setups that circumvent the stringent environmental isolating requirements associated with perfect-crystal neutron interferometry~\cite{rauch2015neutron,saggu2016decoupling} and are capable of working in the full field of the neutron beam~\cite{Clauser1994,Pfeiffer2006,croninnano,daviddifferential, Cronin_2009_RMP,chapman1995near,busi2023multi}. The backbone of these setups is the near-field phenomena of self-imaging known as the ``Talbot effect''~\cite{TalbotEffect}. Further developments introduced phase-grating moir\'e interferometers (PGMIs) that are composed of exclusively phase-gratings and manifest interference patterns that are directly detectable via typical neutron camera~\cite{miao2016universal,pushin2017far,hussey2016demonstration,sarenac2018three,brooks2018neutron,brooks2017neutron}. 

Here we take the next step in the development of the neutron PGMI toolbox by expanding to two-dimensional (2D) moir\'e interference. We explore the role of phase singularities that materials possessing helical structures are expected to induce in the moir\'e pattern and we describe the additional metrics for quantification. Furthermore, we also demonstrate 2D moir\'e interference with orthogonal directionality. The addition of an interference pattern serving as in-situ reference enables novel approaches for high precision measurements of fundamental forces such as the Newtonian constant of gravitation.

\begin{figure*}
\centering\includegraphics[width=.95\linewidth]{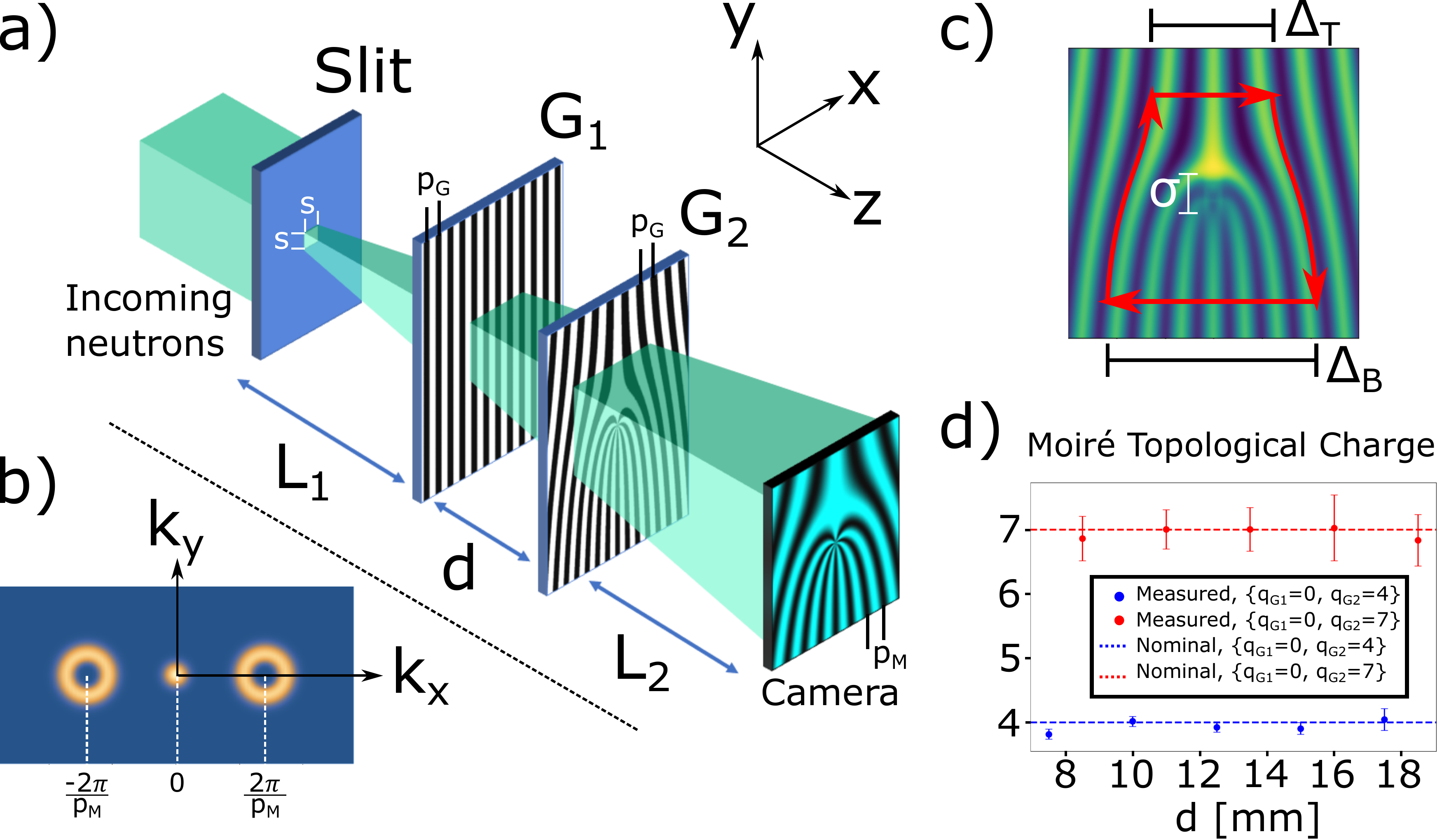}
\caption{a) The two-PGMI setup with fork-dislocation phase-gratings. In our experiments we employed and characterized $\{q_{G_1}=0,q_{G_2}=4\}$ and $\{q_{G_1}=0,q_{G_2}=7\}$ configurations. The moir\'e pattern at the camera manifests a phase-singularity with a topological charge of $q_M=\Delta q_{G}$ and with moir\'e period of $p_M=Lp_G/d$, where $L$ is the distance from the slit to camera, $p_G$ is the period of the phase-gratings, and $d$ is the distance between the two phase-gratings.  b) The Fourier transform of the intensity pattern possesses doughnut-shaped diffraction orders indicative of helical phase structures~\cite{sarenac2022experimental}. c) Simulated intensity profile where the size of the smear $(\sigma)$ centered on the phase-singularity is determined by the size of the slit $(s)$. A convenient method of determining the topology in the measured intensity profile is given by $q_M=N_B-N_T$ where $N_B(N_T)$ is the number of periods in a segment $\Delta_B(\Delta_T)$ below (above) the origin where the connecting vertical lines (red arrows) between the two segments follow a trajectory of constant intensity~\cite{nye1974dislocations}. d) The fitted topological charge of the observed moir\'e interference in our experiments for the $\{q_{G_1}=0,q_{G_2}=4\}$ and $\{q_{G_1}=0,q_{G_2}=7\}$ configurations. The uncertainties shown are statistical. }
 \label{fig:fig1}
\end{figure*}

\begin{figure*}
\centering\includegraphics[width=.95\linewidth]{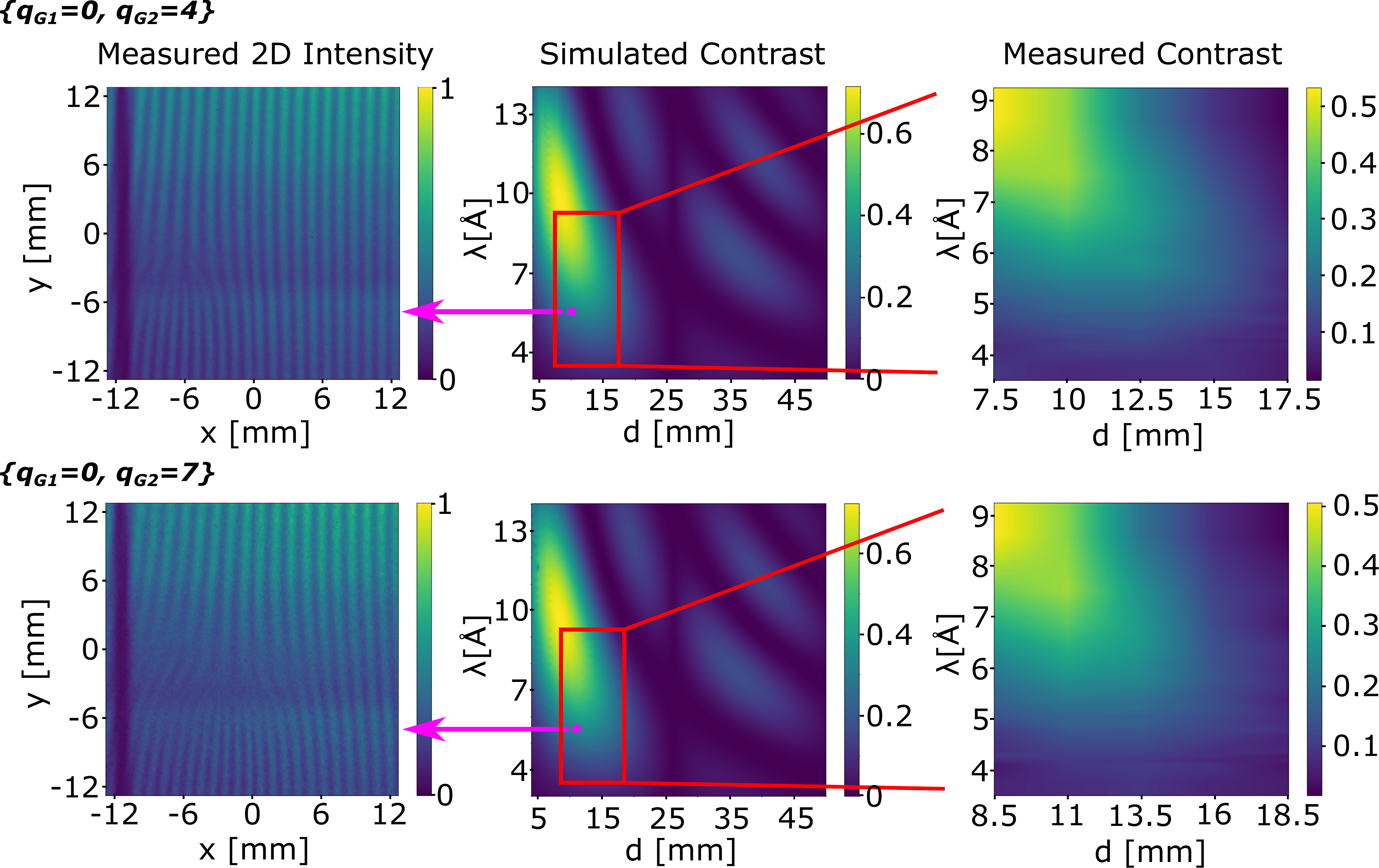}
\caption{The intensity and contrast behaviour of the two-PGMI setup with fork-dislocation phase-gratings. (First column) The observed 2D moir\'e intensity profile at the camera. (Second column) The simulated contrast as a function of wavelength and phase-grating separation distance. (Third column) The measured contrast for the accessible setup parameters. Top (bottom) row shows the setup for the $\{q_{G_1}=0,q_{G_2}=4\}$ ($\{q_{G_1}=0,q_{G_2}=7\}$) configuration. The 2D moir\'e pattern at the camera manifests a phase-singularity with topology $q_M=\Delta q_{G}$ as described in Fig.~\ref{fig:fig1}. The middle column depicts the contrast behaviour for a larger range of parameters, and indicates the parameters for the other two columns. The wavelength range for the the 2D moir\'e intensity profiles shown in the first column is $\lambda=5$~\AA$-6$~\AA~and the phase-grating separation is $d=10~mm~(11~mm)$ for the $\{q_{G_1}=0,q_{G_2}=4\}$ ($\{q_{G_1}=0,q_{G_2}=7\}$) configuration. 
}
 \label{fig:fig2}
\end{figure*}

\begin{figure*}
\centering\includegraphics[width=.95\linewidth]{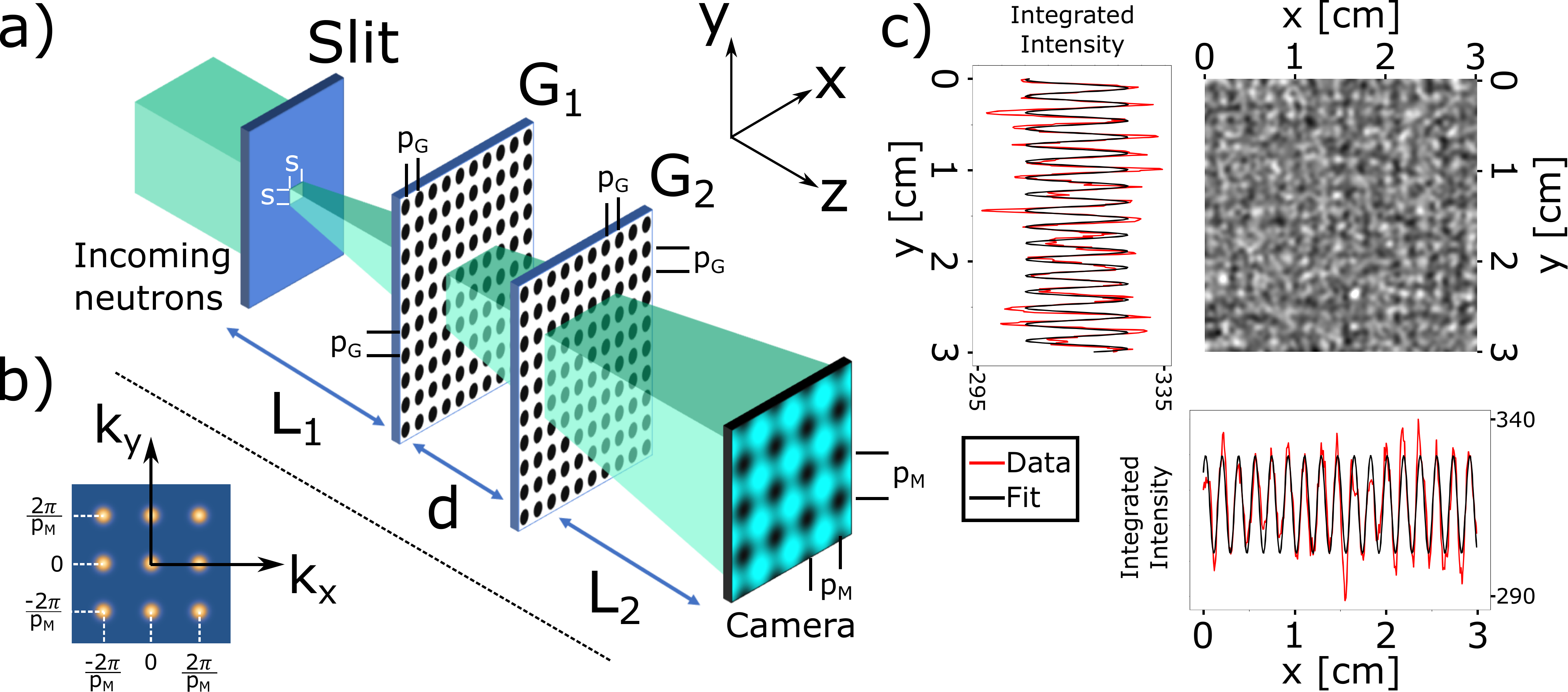}
\caption{a) The two-PGMI setup with two 2D phase-gratings. The moir\'e pattern at the camera manifests a 2D moir\'e pattern that possesses an x and y periodicity of $p_M=Lp_G/d$, where $L$ is the distance from the slit to camera, $p_G$ is the period of the phase-gratings, and $d$ is the distance between the two phase-gratings.  b) The Fourier transform of the intensity pattern shows 2D diffraction orders. c) The observed intensity profile at the camera can be integrated along x or y to independently analyze the contrast along the two orthogonal directions, thereby providing an in-situ reference signal when considering 1D forces and structures. 
}
 \label{fig:fig3}
\end{figure*}

\section{\label{sec:level1}Methods}

We fabricated four types of phase-gratings: 2D phase-gratings, and fork-dislocation phase-gratings with topological charge of $q=0,4,7$. Note that the $q=0$ is a typical 1D phase-grating. The detailed nanofabrication procedure and scanning electron microscopy (SEM) images of all four cases can be found in Appendix A. All phase-gratings were fabricated out of silicon and had a target period of $3~\mu m$ and height of $8.53~\mu m$. The height of $8.53~\mu m$ was chosen so that the grating imparts an optimal $\pi/2$ phase shift for $\lambda=9$~\AA~neutrons.

The experiments with fork-dislocation phase-gratings were performed at the ASTERIX facility at the Los Alamos Neutron Science Center (LANSCE)~\cite{nelson2018neutron}. The wavelength distribution is shown in Appendix B. The slit was $500~\mu m$ by $500~\mu m$ in size and the distance from the slit to the first phase-grating (camera) was $2.13~m$ $(4.25~m)$. The camera pixel size was $50~\mu m$ by $50~\mu m$. The data acquisition time was $\approx 20$ min at each grating separation distance. A detector normalization image was obtained by summing phase stepping measurements.

The experiments with the 2D phase-gratings were performed at the the RADEN facility at the Japan Proton Accelerator Research Complex (J-PARC)~\cite{shinohara2020energy}.  The wavelength distribution is shown in Appendix B. The 2D slit was composed of a sequence of two perpendicular 1D slits each made by bringing two cadmium pieces together with the target gap of $500~\mu m$. Two experimental setups were used, one for polychromatic measurements and the other with a neutron camera with lower dark counts ($\approx 0$) for time-of-flight resolved measurements. For the first configuration the distance from the slit to the first phase-grating (camera) was $4.16~m$ $(8.30~m)$, the camera pixel size was $100~\mu m$ by $100~\mu m$, and the image acquisition time was 4~h at each grating separation distance. For the second configuration the distance from the slit to the first phase-grating (camera) was $4.23~m$ $(8.53~m)$, the camera pixel size was $31~\mu m$ by $31~\mu m$, and the data acquisition time was 5~h. For each setup configuration a detector normalization image was obtained from a measurement at a setup configuration that results in zero contrast. Furthermore, due to the presence of significant salt and pepper noise a median filter of 5 x 5 pixels was applied to the final images.

\section{\label{sec:level1}Results and Discussion}

\subsection{Fork-dislocation phase-gratings}

Here we explore the effect of helical structures that manifest phase singularities in the moir\'e pattern by introducing a topological charge onto the phase-gratings themselves. A fork-dislocation phase-grating with period $p_G$, height $D$, and topological charge $q_G$ has the profile:


\begin{align}
\Phi=\frac{Nb_c\lambda D}{2}\text{sgn}\left[\cos\left(k_G x+q_G\phi\right)\right]
\label{Eqn:forkgrating}
\end{align}
\noindent where $k_G=2\pi/p_G$ is the grating wave vector, $x(\phi)$ is the Cartesian (azimuthal) coordinate,  $Nb_c$ is the scattering length density of the grating material, and $\lambda$ is the neutron wavelength. Using the recently introduced k-space model~\cite{sarenac2023cone} we can simulate the intensity and contrast behaviour using the phase-grating profile of Eq.~\ref{Eqn:forkgrating}.

An experimental demonstration was performed with the two-PGMI configurations of $\{q_{G_1}=0,q_{G_2}=4\}$ and $\{q_{G_1}=0,q_{G_2}=7\}$. The first configuration is depicted on Fig.~\ref{fig:fig1}a. The moir\'e pattern at the camera possesses a topological charge of $q_M=\Delta q_{G}$ with moir\'e period of $p_M=Lp_G/d$, where $L$ is the distance from the slit to camera, $p_G$ is the period of the phase-gratings, and $d$ is the distance between the two phase-gratings. The Fourier transform of the intensity profile is shown in Fig.~\ref{fig:fig1}b where the doughnut profiles are indicative of helical structures with phase singularities~\cite{sarenac2022experimental}.  

Whereas contrast is the figure of merit for a 1D two-PGMI, an additional metric is needed for identifying and characterizing phase singularities. In this particular setup the topology in the moir\'e pattern can be determined by $q_M=N_B-N_T$ where $N_B(N_T)$ is the number of periods in a segment $\Delta_B(\Delta_T)$ below (above) the origin where the connecting vertical lines between the two segments follow a trajectory of constant intensity~\cite{nye1974dislocations}, see Fig.~\ref{fig:fig1}c. Using this method we can calculate the measured moir\'e topology using a fit procedure for the two experimental configurations as shown in Fig.~\ref{fig:fig1}d. This method becomes increasingly useful as the smear centered on the phase-singularity increases with slit size. 

Examples of the measured moir\'e pattern at the camera are shown in the first column of Fig.~\ref{fig:fig2} for both $\{q_{G_1}=0,q_{G_2}=4\}$ and $\{q_{G_1}=0,q_{G_2}=7\}$. Note that the vertical stripe on the left is an upstream optical effect due to the guide geometry, and unrelated to the moir\'e interference. The simulated (measured) contrast as a function of phase-grating separation and wavelength is shown in the second (third) column of Fig.~\ref{fig:fig2}. The experimentally accessible parameters were not centered around the optimal contrast parameters as the phase-gratings were initially designed for a $\lambda=9$~\AA~neutrons. Good agreement is found between the predicted and measured contrast.

\subsection{2D phase-gratings}

Here we aim to explore the 2D moir\'e pattern that possesses a periodicity along two orthogonal directions. There are many variations possible for 2D phase-grating profiles. For our experiments we chose a profile of:


\begin{align}
\Phi=\frac{Nb_c\lambda D}{2}\text{sgn}\left[\cos\left(k_G x\right)+\cos\left(k_G y\right)-1\right]
\label{Eqn:2Dgrating}
\end{align}
\noindent which is essentially a 2D array of circular holes as shown in the SEM images of Appendix A. The setup schematic is depicted on Fig.~\ref{fig:fig3}a. The 2D moir\'e pattern at the camera possesses a sinusoidal pattern in both the x and y directions. Therefore, its Fourier transform shows 2D diffraction orders as depicted in Fig.~\ref{fig:fig3}b. The 2D moir\'e pattern can be integrated along either Cartesian direction as shown in Fig.~\ref{fig:fig3}c where we consider the y-axis to be along Earth's gravity and the x-axis along the perpendicular direction. The shown profile is obtained by considering the phase-grating separation of $d=12.5~mm$ and $\lambda=5$~\AA$-6$~\AA~wavelength distribution, and the observed intensity profile at the camera has been rotated by $\theta=13.3\degree$ (corresponding to the maximum contrast location in a contrast vs $\theta$ plot).

\begin{figure*}
\centering\includegraphics[width=.95\linewidth]{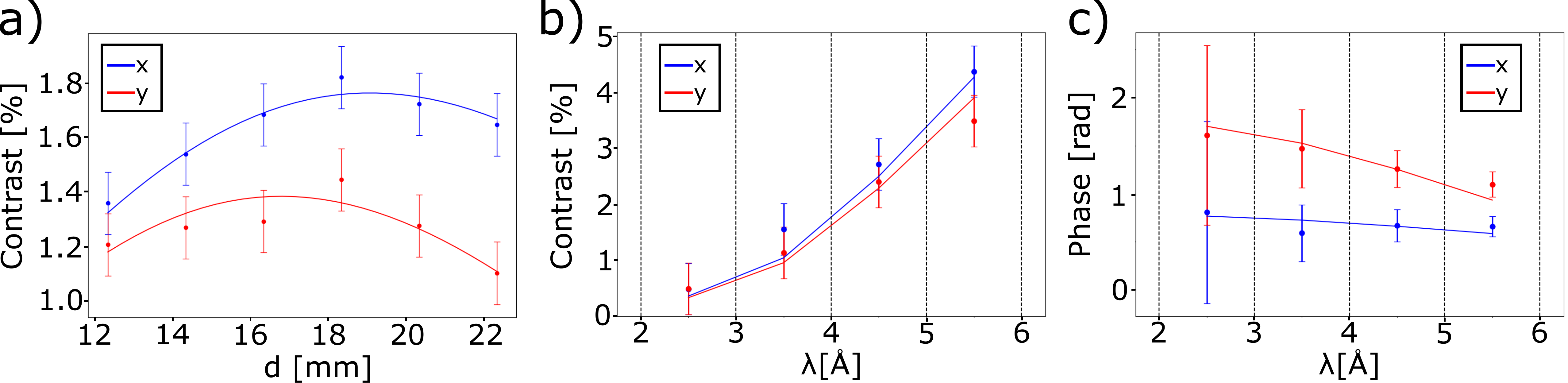}
\caption{a) The measured contrast of the two-PGMI setup with two 2D phase-gratings as a function of grating separation distance $d$ and considering the full polychromatic wavelength distribution (see Appendix B). The simulation curves are obtained with the k-space model~\cite{sarenac2023cone} for the given experimental parameters and with a least squares fit for the two slit sizes. The best fit is determined for $643~\mu$m $\pm49~\mu$m for the slit along x and $790~\mu$m $\pm50~\mu$m for the slit along y, which is well within the expected experimental error. b) The wavelength-dependent contrast along the x and y direction for $d=12.5~mm$. c) The wavelength-dependent phase shift along the x and y direction for $d=12.5~mm$. The wavelength distribution around each wavelength interval is shown in Appendix B. We find good agreement with the expected values when taking into account the gravitational fall from the second phase-grating to the camera and the relative rotation of $\theta=13.3\degree$ between the moir\'e vector and the Earth's gravitational vector. The uncertainties shown are statistical. 
}
 \label{fig:fig4}
\end{figure*}

The measured contrast as a function of phase-grating separation ($d$) for a polychromatic wavelength distribution is shown on Fig.~\ref{fig:fig4}a, and Fig.~\ref{fig:fig4}b shows wavelength-dependent contrast for a particular $d$. See Appendix B for the wavelength distribution profile. Using the k-space model of Ref.~\cite{sarenac2023cone} it can be confirmed that even when accounting for gravity the difference between the contrasts along the two directions should have been negligible for the given experimental parameters, and that the observed difference is most likely due to the difference in slit sizes. The 2D slit was composed of a sequence of two perpendicular 1D slits each made by bringing two cadmium pieces together with the target gap of $500~\mu m$. Performing a least squares fit to the two slit sizes we find good agreement with $643~\mu$m $\pm49~\mu$m for slit along x and $790~\mu$m $\pm50~\mu$m for slit along y. These values are well within the expected experimental uncertainties. 

Fig.~\ref{fig:fig4}c shows the wavelength-dependent phase shift that can be used to quantify the effect of gravity. 
In a two-PGMI the main contribution to the gravitationally induced phase shifts is the neutron fall $(\Delta_y)$ between the second phase-grating and the neutron camera: 

\begin{align}
\phi_g=\frac{2\pi\Delta_y}{p_M}\cos\theta+C_0=\frac{\pi g}{p_M}\left(\frac{L_2 m \lambda}{h}\right)^2\cos\theta+C_0
\label{Eqn:gravitypahse}
\end{align}

\noindent where $\theta$ is the angle between the moir\'e vector and the gravitational force vector, $g$ is the acceleration due to Earth's gravity, $m$ is the mass of the neutron, $h$ is the Plank's constant, and $C_0$ is an arbitrary offset. 
We find good agreement between the expected and observed wavelength-dependent phase shifts.

\section{Conclusion}

We have expanded neutron phase-grating moir\'e interferometry to 2D and enabled the use of new degrees of freedom for material characterization studies and high precision measurements of fundamental constants. We examined the manifestation and characterization of phase singularities in the moir\'e pattern by incorporating fork-dislocation phase-gratings. Future studies will look at the interference between multiple phase singularities and the effects that would be introduced by samples with phase singularities such as skyrmions~\cite{milde2013unwinding,Yu2020real}. We also characterized two-PGMI setups that simultaneously manifest moir\'e interference along two orthogonal directions. The orthogonal interference pattern enables the presence of an in-situ reference signal that can greatly reduce systematic errors. Furthermore, future studies will also examine the use of 2D phase-linear gratings with three-PGMI where it is possible to substantially increase the distance between the phase-gratings. Whereas Eq.~\ref{Eqn:gravitypahse} is considering the neutron gravitation fall relative to the moir\'e period at the camera, a three-PGMI has the capability to consider the neutron gravitation fall relative to the phase-grating period: $\phi_g\propto2\pi\Delta_y/p_G$. This can provide an amplification to phase sensitivity by several orders of magnitude.

\section*{Acknowledgements}

This work was supported by the Canadian Excellence Research Chairs (CERC) program, the Natural Sciences and Engineering Research Council of Canada (NSERC) Discovery program, Collaborative Research and Training Experience (CREATE) program, the Canada  First  Research  Excellence  Fund  (CFREF), and the National Institute of Standards and Technology (NIST) and the US Department of Energy, Office of Nuclear Physics, under Interagency Agreement 89243019SSC000025. The pulsed neutron experiment at J-PARC MLF was performed
under a user program (Proposal No. 2022A0104).

\bibliography{OAM}

\begin{thebibliography}{39}
\bibitem [{NIS()}]{NIST_disclamer}%
  \BibitemOpen
  \href@noop {} {}\bibinfo {note} {Certain trade names and company products are
  mentioned in the text or identified in an illustration in order to adequately
  specify the experimental procedure and equipment used. In no case does such
  identification imply recommendation or endorsement by the National Institute
  of Standards and Technology, nor does it imply that the products are
  necessarily the best available for the purpose.}\BibitemShut {Stop}%
\end{thebibliography}

\clearpage
\onecolumngrid

\section*{Appendix}

\subsection{Phase-grating fabrication}

\begin{figure*}[!h]
\centering\includegraphics[width=\linewidth]{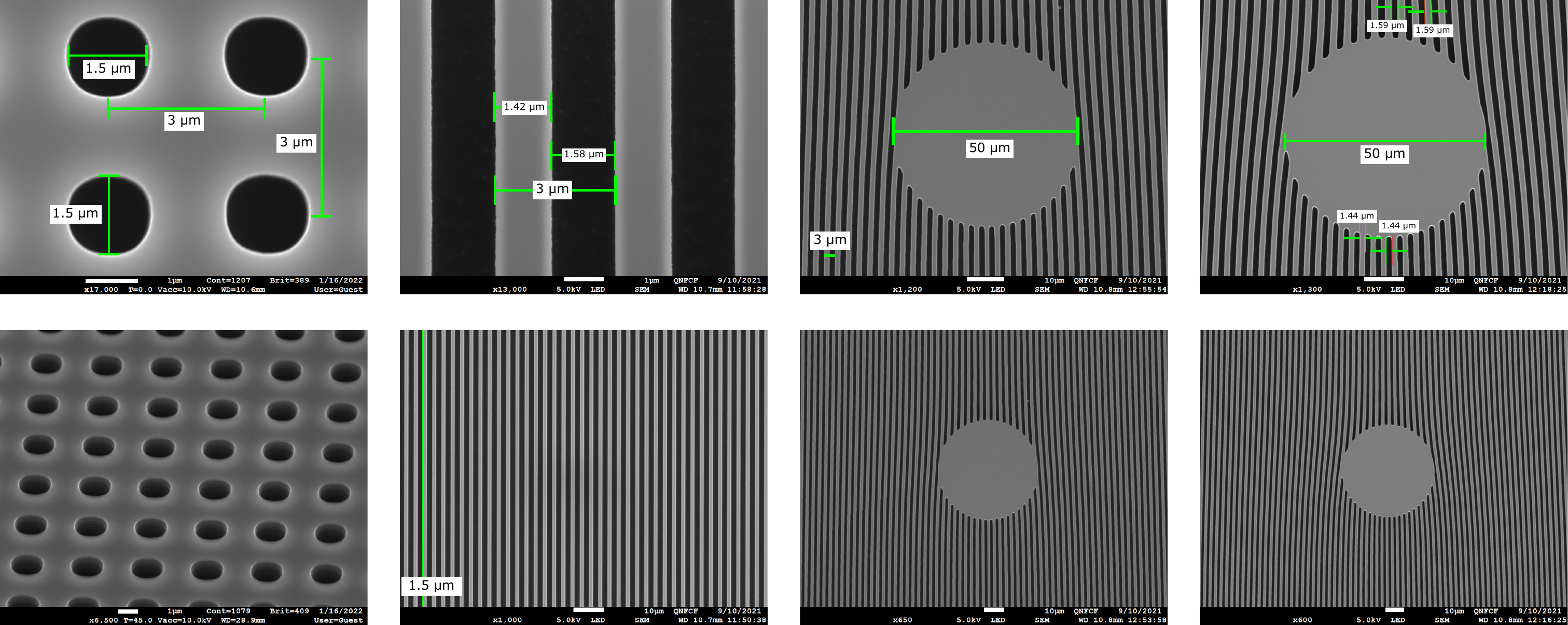}
\caption{SEM profiles of the phase-gratings used in the experiments. All phase-gratings were fabricated out of silicon and had a target period of $3~\mu m$ and height $8.53~\mu m$. The first column depicts the 2D phase-gratings, the second, third, fourth column shows the fork-dislocation phase-grating with $q=0,4,7$, respectively. The $q=0$ are the typical 1D phases-gratings. There was a $50~\mu m$ mask covering the middle region of the  $q=4$ and $q=7$ fork-dislocation phase-gratings. This is a common practice to avoid the fabrication challenges associated with the higher aspect ratio near the phase-singularity. In regards to the presented PGMI configurations, the masked region sets the limit on the resolution of the observable moir\'e phase-singularity to $\approx100~\mu$m. Therefore the effect of this masked region is negligible. }
 \label{fig:SEM}
\end{figure*}

Double-side polished 10.16~cm diameter (100) silicon wafers were used to fabricate these gratings. A bilayer resist (PMGI/S1805~\cite{NIST_disclamer}) was patterned via a maskless aligner (MLA 150, Heidelberg Instrument). As a hard mask for the plasma etching, Cr metal (60 nm) was e-beam evaporated and lifted-off in heated PG Remover. A Bosch recipe was adopted to achieve a vertical sidewall etch profile. The samples were etched in an Oxford PlasmaLab ICP-380 inductively coupled plasma reactive ion etching (ICP-RIE) system, which provides high-density plasma with independently controlled system parameters. In our Bosch recipe, the passivation half cycle comprises the RF chuck power: 5 W, ICP coil power: 1000 W, C4F8: 160 sccm, pressure: 2.67 Pa, temperature: 15 $\degree$C for 5 s while the etch half cycle comprises the RF chuck power: 100 W, ICP coil power:~1000 W, SF6: 160 sccm, pressure: 3.33 Pa, temperature: 15 $\degree$C for 4 s. After fabricating the gratings, the remaining Cr mask was removed via plasma etching. Fig.~\ref{fig:SEM} shows the SEM images of the four types of phase-gratings that were fabricated.

\clearpage
\subsection{Wavelength Distributions}

\begin{figure*}[!h]
\centering\includegraphics[width=\linewidth]{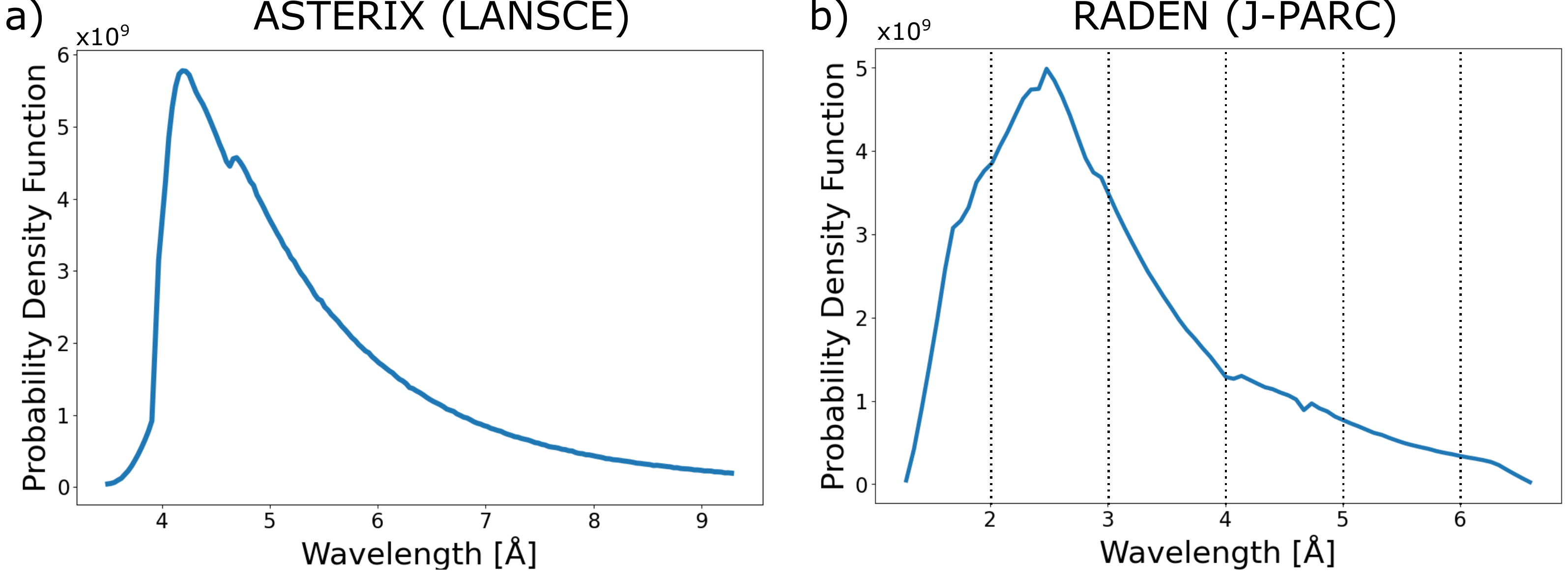}
\caption{a) The wavelength distribution for the ASTERIX facility at the Los Alamos Neutron Science Center (LANSCE). This wavelength distribution was used for the data presented in Fig.~\ref{fig:fig1}d and Fig.~\ref{fig:fig2}. b) The wavelength distribution for the RADEN facility at the Japan Proton Accelerator Research Complex (J-PARC). This wavelength distribution was used for the data presented in Fig.~\ref{fig:fig3}c and Fig.~\ref{fig:fig4}. The vertical lines show the 4 regions of the time-of-flight data used in Figs.~\ref{fig:fig4}b~\&~\ref{fig:fig4}c.}
 \label{fig:wavelength}
\end{figure*}

\end{document}